\documentclass[english,twocolumn,nofootinbib]{revtex4}
\usepackage[latin1]{inputenc}
\usepackage{array}
\usepackage{longtable}
\usepackage{amsmath}
\usepackage{graphicx}
\usepackage{amssymb}

\makeatletter

\providecommand{\tabularnewline}{\\}

\renewcommand{\vec}[1]{\mathbf{#1}}

\renewcommand*{\sectionmark}[1]{} 
\renewcommand*{\subsectionmark}[1]{} 

\newcommand{\refcite}[1]{Ref.~\onlinecite{#1}}

\usepackage{babel}
\makeatother
\begin{document}

\title{Out-of-plane spin polarization from in-plane electric and magnetic
fields}

\author{Hans-Andreas Engel, Emmanuel I. Rashba, and Bertrand I. Halperin}

\affiliation{Department of Physics, Harvard University, Cambridge, Massachusetts
02138 }

\begin{abstract}
We show that the joint effect of spin-orbit and magnetic fields leads
to a  spin polarization perpendicular  to the plane of a two-dimensional
electron system with Rashba spin-orbit coupling and in-plane parallel
dc magnetic and electric fields, for angle-dependent impurity scattering
or nonparabolic energy spectrum,  while only in-plane polarization
persists for simplified models. We derive Bloch equations,  describing
the main features of recent experiments, including the magnetic field
dependence of static and dynamic responses.


\end{abstract}
\maketitle

\newcommand{\ZeemanX}{B}

\newcommand{\azimuthalAngle}{\varphi}

\renewcommand{\ZeemanX}{\Delta_x}

\newcommand*{\ZeemanY}{\Delta_y}

\newcommand{\ket}[1]{|#1\rangle}

\newcommand{\bra}[1]{\langle#1|}

\newcommand{\braopket}[3]{\left\langle #1\left|#2\right|#3\right\rangle }

\newcommand{\braket}[2]{\langle#1|#2\rangle}

\newcommand{\bsigma}{\boldsymbol{\sigma}}

\newcommand{\brho}{\boldsymbol{\rho}}

\newcommand{\bzeta}{\boldsymbol{\zeta}}

\newcommand{\bvarepsilon}{\boldsymbol{\varepsilon}}

\newcommand{\bGamma}{\boldsymbol{\Gamma}}

\newcommand{\bkappa}{\boldsymbol{\kappa}}

\newcommand{\btau}{\boldsymbol{\tau}}

\newcommand{\bpi}{\boldsymbol{\pi}}

\newcommand{\dotvec}[1]{\vec{\dot{#1}}}

\newcommand{\lambdaCBar}{\lambdabar_{\mathrm{c}}}

\newcommand{\ShermanBmAvg}{\gamma}

\newcommand{\PWinkler}{P_{\mathrm{W}}}

\newcommand{\PNozieres}{P_{\mathrm{NL}}}

\newcommand{\PAndrada}{P_{\mathrm{AndradaDS}}}

\newcommand{\currentParticle}[2]{j_{#2,\,\mathrm{p}}^{#1}}

\newcommand{\jSH}{\vec{j}}

\newcommand{\SOcoupling}{\lambda}

\newcommand{\HSO}{H_{\mathrm{SO}}}

\newcommand{\polAHE}{\vec{P}_{\mathrm{AH}}}

\newcommand{\jSHc}{j}

\newcommand{\jSHSJ}{\jSH_{\,\mathrm{SJ}}}

\newcommand{\SOsplitting}{\Delta_{0}}

\newcommand{\Egap}{E_{\mathrm{0}}}

\newcommand{\polDensity}{\vec{s}^{E}}

\newcommand{\polDensityElem}{s^{E}}

\newcommand{\polDensityElemDot}{\dot{s}^{E}}

\newcommand{\polDensityIso}{\vec{s}^{\lblIso}}

\newcommand{\polDensityAniso}{\vec{s}^{\lblAniso}}

\newcommand{\polDensityElemAniso}{s^{\lblAniso}}

\newcommand{\polDensityElemDotAniso}{\dot{s}^{\lblAniso}}

\newcommand{\lblAniso}{\mathrm{ai}}

\newcommand{\polDensityEq}{\vec{s}^{\mathrm{eq}}}

\newcommand{\polDensityElemEq}{s^{\mathrm{eq}}}

\newcommand{\polDensityElemDotEq}{\dot{s}^{\mathrm{eq}}}

\newcommand{\polDensityTot}{\vec{s}}

\newcommand{\polDensityElemTot}{s}

\newcommand{\polDensityElemDotTot}{\dot{s}}

\newcommand{\polDensityDotTot}{\vec{\dot{s}}}

\newcommand{\jSpin}[2]{j_{#2}^{#1}}

\newcommand{\jChargeComp}[1]{j_{#1}^{\mathrm{c}}}

\newcommand{\energyOfK}[1]{\epsilon_{#1}}

\newcommand{\nTwoD}{n_{2\mathrm{D}}}

\newcommand{\DOS}{\nu}

\newcommand{\tauSpinInPlane}{\tau_{xy}}

\newcommand{\tauSpinz}{\tau_{z}}

\newcommand{\tauSpinGeomMean}{\tau_{\mathrm{s}}}

\newcommand{\invTauSpinTensor}{{\overleftrightarrow{\tau}\hspace{-.5mm}{}_{\mathrm{s}}^{-1}}}

\newcommand{\fDiagonal}{f_{\mathrm{c}}}

\newcommand{\fSpin}{\vec{f}}

\newcommand{\velocityE}{v_{\epsilon}}

\newcommand{\Vimpurity}{V_{\mathrm{i}}}

\newcommand{\fMatrixE}{\hat{\Phi}}

\newcommand{\fSpinE}{\boldsymbol{\Phi}}

\newcommand{\fSpinEComponent}[1]{\Phi^{#1}}

\newcommand{\fDiagonalE}{n}
 
\newcommand{\Eparam}{\beta}
 
\newcommand{\PhiPM}{\Psi}

\newcommand{\PhiPMm}[1]{\PhiPM_{#1}}

\newcommand{\PhiPMdotm}[1]{\dot{\PhiPM}_{#1}}

\newcommand{\Phimum}[2]{\Phi_{#2}^{#1}}

\newcommand{\Phizm}[1]{\Phimum{z}{#1}}

\newcommand{\Phixm}[1]{\Phi_{#1}^{x}}

\newcommand{\Phiym}[1]{\Phi_{#1}^{y}}

\newcommand{\Phiydotm}[1]{\dot{\Phi}_{#1}^{y}}

\newcommand{\Phizdotm}[1]{\dot{\Phi}_{#1}^{z}}

\newcommand{\driftField}{\vec{b}_{\mathrm{dr}}}

\newcommand{\driftFieldy}{b_{\mathrm{dr}}^{y}}

\newcommand{\bIntrinsicVec}{\vec{b}^{\left(\mathrm{i}\right)}}

\newcommand{\scatteringAngle}{\gamma}

\newcommand{\gammaPlusTildeN}{\tilde{\gamma}_{\left(+\right)}}

\newcommand{\gammaMinusTildeN}{\tilde{\gamma}_{\left(-\right)}}

\newcommand{\gammaPlusMinusTildeN}{\tilde{\gamma}_{\left(\pm\right)}}

\newcommand{\gammaHomTildeN}{\tilde{\gamma}_{0}}

\newcommand{\bRashba}{b_{\alpha}}

\newcommand{\bDresselhausLinear}{b_{\beta}}

\newcommand{\DHbeta}{\beta_{\mathrm{D}}}

\newcommand{\lblIso}{\mathrm{is}}

\newcommand{\fSpinEisoSol}{\fSpinE_{\lblIso}}

\newcommand{\anisoPrefix}{\delta}

\newcommand{\fSpinEanisoCorr}{\fSpinE_{\lblAniso}}

\newcommand{\bAvgy}{\left\langle b_{y}\right\rangle }

\newcommand{\kUnitVec}{\hat{\vec{k}}}

Generating spin populations at a nanometer scale is one of the central
goals of spintronics \cite{Wolf_Spintronics}. Using spin-orbit interaction
promises electrical control, allowing to integrate spin generation
and manipulation into the traditional architecture of electronic devices.
Bulk spin polarization, driven by electron drift in an electric field,
was predicted long ago for noncentrosymmetric three- (3D) and two-dimensional
(2D) systems \cite{IvchenkoPikusSpinPol,VaskoPrima79,LNMSpinPolarization,Edelstein90,ALGPspinPol91,BernevigStrain}.
In 2D, the polarization is in-plane, typically along the effective
spin-orbit field $\driftField=\langle\vec{b}_{\mathrm{SO}}(\vec{k})\rangle\neq0$,
obtained by averaging spin-orbit coupling over the distribution of
electron momenta $\hbar\vec{k}$~\cite{footnoteAnisobdr}. In-plane
polarization components were observed recently in  $p$-GaAs heterojunctions
\cite{SilovBW04}, quantum wells \cite{GanichevPolarization}, and
strained $n$-InGaAs films \cite{KatoBulkPolarization04}. Out-of-plane
spin polarization can be generated by the spin-Hall effect, but only
near sample edges \cite{DPpolarization}. Below, we propose a mechanism
for out-of-plane spin polarization generated in the bulk by applying
an in-plane magnetic field $\vec{B}$. This perpendicular polarization
allows efficient optical access, e.g., via Kerr rotation.  We find
that the use of such an average field $\driftField$ is not always
valid. Naively, one might consider the system as being subject to
a total in-plane field $\langle\vec{b}\rangle$, given by the sum
of $\vec{B}$ and $\driftField$, see Fig.~1(a). In steady state,
one then expects electrons to be polarized along this total field:
in particular, no polarization perpendicular to the $\driftField,\:\vec{B}$
plane. Algebraic addition of these fields worked well in describing
Hanle precession of optically oriented 2D electrons in GaAs \cite{JetpLett_52_230}.
However, Kato et al.\ \cite{KatoBulkPolarization04} reported a spin
polarization that is incompatible with such a naive picture and emphasized
the need of identifying its microscopic mechanisms. 

\begin{figure}
\begin{center}\includegraphics[%
  width=87mm,
  keepaspectratio]{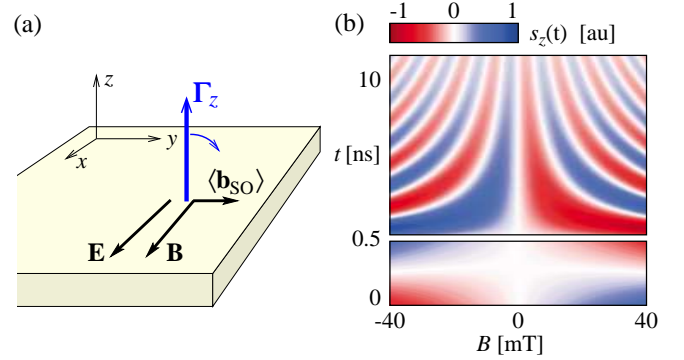}\end{center}

\caption{(color) (a) Field geometry, assuming $g\mu_{\mathrm{B}}>0,$ $\alpha<0$.
Out-of-plane spin polarization is electrically generated with rate
$\Gamma_{z}$ (blue arrow) due to the interplay between spin-orbit
interaction, external electric field $\vec{E}$ and magnetic field
$\vec{B}$, and anisotropic impurity scattering. The polarization
precesses (blue arc) in $g^{*}\mu_{\mathrm{B}}\vec{B}+\langle\vec{b}_{\mathrm{SO}}\rangle$.
(b) Dynamics of out-of-plane component of spin polarization generated
by a short electrical pulse of length $t_{p}\lesssim1\:\mathrm{ns}$,
for $\gammaHomTildeN=0.03$, $g^{*}=0.65$ and $\tauSpinz=5\:\mathrm{ns}$.
This pattern of spin polarization is in agreement with the experimental
data of Fig.~4(c), \refcite{KatoBulkPolarization04}.}
\end{figure}

In this article, we develop a theory describing the interplay between
spin-orbit interaction and external electric and magnetic fields in
the presence of impurity scattering, and demonstrate that the concept
of average spin-orbit field is subject to severe restrictions. The
naive expectation turns out to be correct only in the special case
of parabolic bands and isotropic impurity scattering.  However, as
we show below, for anisotropic scattering (e.g., small angle scattering),
such correlations result in a more complex structure of the distribution
function and an out-of-plane spin polarization. Concretely, \emph{interplay
of $\driftField$ and $\vec{B}$ leads to a generation term in the
Bloch equation proportional to} $\driftField\times\vec{B}$ whose
magnitude is controlled by anisotropy of potential scattering and
non-parabolicity of the energy spectrum\emph{.} Remarkably, while
this does not change the symmetry of the Hamiltonian, the symmetry
of responses is lower than in the special case (and in the naive picture).
Our results give a microscopic explanation of experiments \cite{KatoBulkPolarization04}
and provide a novel mechanism for generating spin polarization electrically
via spin-orbit interaction.

We consider a model of 2D electrons with charge $e<0$ and (pseudo-)
spin $\frac{1}{2}$, obeying a Hamiltonian\begin{equation}
H=\energyOfK{k}-{\textstyle \frac{1}{2}}\vec{b}\left(\vec{k}\right)\cdot\bsigma+V\left(\vec{r}\right),\label{eq:H}\end{equation}
where $\energyOfK{k}$ is the dispersion law in the absence of spin-orbit
coupling, $V\left(\vec{r}\right)$ is the potential due to impurites,
$V_{\mathrm{i}}(\vec{r})$, plus a small electric field $\vec{E}$,
$\bsigma$ are the Pauli spin matrices, and $\vec{b}\left(\vec{k}\right)$
includes both intrinsic spin-orbit field $\vec{b}_{\mathrm{SO}}(\vec{k})$
and external field $\vec{B}$. We consider in-plane magnetic field,
i.e., there is no orbital quantization, and disregard electron-electron
interaction. In the following, we study spin polarization density
$\polDensityTot\left(\vec{r}\right)=\left\langle \bsigma\right\rangle \nTwoD$
and spin currents $\vec{j}^{\mu}\left(\vec{r}\right)=\langle{\textstyle \frac{1}{2}}\,\{\sigma_{\mu},\:\vec{v}\}\rangle\nTwoD$.
Here, $\nTwoD$ is the electron density, $\left\{ \,,\,\right\} $
is the anticommutator, and the velocity $\vec{v}=i\left[H,\,\vec{r}\right]$
is spin-dependent. (We set $\hbar=1$.)

For a bulk 2D system with only intrinsic spin-orbit interaction, the
kinetic equation has been derived~\cite{ALGPspinPol91,Khaetskii_2DGeneralized,ShytovKinetic}.
Following \refcite{ShytovKinetic}, we may write a spin-dependent
Boltzmann equation for the distribution function, represented as a
$2\times2$ spin matrix $\hat{f}=\hat{f}_{0}(\vec{k})+\frac{1}{2}\fDiagonal(\vec{k})\,1\!\!1+\fSpin(\vec{k})\cdot\bsigma$,
with equilibrium distribution function $\hat{f}_{0}$, excess particle
density $\fDiagonal$, $\vec{k}=(k_{x},\: k_{y})=(k\cos\azimuthalAngle,\: k\sin\azimuthalAngle)$,
and spin polarization density described by $\fSpin$. Magnetic field
and spin-orbit coupling split the energy spectrum into two branches:
for a given energy $\epsilon$, there are two Fermi surfaces. Thus,
for elastic scattering, energy $\epsilon$ is conserved but $\left|\vec{k}\right|$
is not, due to inter-branch scattering. In the following, we assume
$b\ll E_{\mathrm{F}}$. This motivates defining $k_{\epsilon}$ such
that $\energyOfK{k_{\epsilon}}=\epsilon$, and defining $v_{\epsilon}=\energyOfK{k_{\epsilon}}'$.
For a fixed energy $\epsilon$, the velocity operator  is \cite{epapsPolarization}\begin{equation}
\vec{v}=\kUnitVec\:\left[v_{\epsilon}+\left(1+\zeta\right)\vec{b}\cdot\bsigma/2k_{\epsilon}\right]-(1/2)\,\partial(\vec{b}\cdot\bsigma)/\partial\vec{k},\label{eq:vnotTovepsilon}\end{equation}
with unit vector $\kUnitVec=\vec{k}/k$ and band nonparabolicity
$\zeta=(k_{\epsilon}/v_{\epsilon})(\partial v_{\epsilon}/\partial k_{\epsilon})-1$.
 Instead of using the distribution function $\hat{f}\left(\vec{k}\right)$
as density in $\vec{k}$-space, we consider it as a function of energy
$\epsilon$ and direction $\azimuthalAngle$ in $\vec{k}$-space.
In this representation, $\fDiagonal\left(\vec{k},\,\epsilon\right)$
and $\fSpin\left(\vec{k},\,\epsilon\right)$ are transformed into
distribution functions $\fDiagonalE\left(\azimuthalAngle,\,\epsilon\right)$
and $\fSpinE\left(\azimuthalAngle,\,\epsilon\right)$, resp., which
can be written as a matrix $\fMatrixE\left(\azimuthalAngle,\,\epsilon\right)$;
for a detailed derivation see \refcite{ShytovKinetic}.   The kinetic
equation  for $\vec{E}=E\vec{\hat{x}}$ is \cite{ShytovKinetic}
\begin{gather}
\frac{\partial\fMatrixE}{\partial t}+\bsigma\cdot\left[\vec{b}\times\fSpinE-\frac{\fDiagonalE}{4v_{\epsilon}}\,\vec{b}\times\frac{\partial\vec{b}}{\partial k}\right]+\frac{eE}{\left(2\pi\right)^{2}}\,\frac{\partial f_{0}}{\partial\epsilon}\nonumber \\
\times\bigg[k_{x}+\frac{1}{2\velocityE}\frac{\partial}{\partial\azimuthalAngle}\left(\vec{b}\cdot\bsigma\,\sin\azimuthalAngle\right)\bigg]={\left(\frac{\partial\fMatrixE}{\partial t}\right)\negthickspace,\negmedspace\negthickspace}_{\mathrm{coll.}}\label{eq:BlhsIntrinsic}\end{gather}
where $f_{0}$ is the Fermi distribution function, $\vec{b}=\vec{b}(\azimuthalAngle)$
is evaluated for $\left|\vec{k}\right|=k_{\epsilon}$, and with charge
distribution $\fDiagonalE=8\Eparam\tau v_{\epsilon}k_{\epsilon}\cos\azimuthalAngle$
and $\Eparam=(eE/16\pi^{2}v_{\epsilon})\,(-\partial f_{0}/\partial\epsilon)$.
In Eq.~(\ref{eq:BlhsIntrinsic}), the first term is the partial time-derivative,
the second term describes spin precession in the momentum dependent
field $\vec{b}\left(\azimuthalAngle\right)$, and the third term is
the driving term, given in lowest order in $\vec{E}$. 

 The collision integral on the r.h.s.\ of Eq.~(\ref{eq:BlhsIntrinsic})
can be found in Born approximation by Golden Rule~\cite{ShytovKinetic},
\begin{gather}
\left(\frac{\partial\fMatrixE(\azimuthalAngle)}{\partial t}\right)_{\mathrm{coll}}=\int_{0}^{2\pi}d\azimuthalAngle'\: K(\vartheta)\left[\fMatrixE(\azimuthalAngle')-\fMatrixE(\azimuthalAngle)\right]\nonumber \\
+\int_{0}^{2\pi}d\azimuthalAngle'\:\bsigma\cdot\left[\vec{M}\left(\azimuthalAngle,\azimuthalAngle'\right)\fDiagonalE\left(\azimuthalAngle'\right)-\vec{M}\left(\azimuthalAngle',\azimuthalAngle\right)\fDiagonalE\left(\azimuthalAngle\right)\right].\label{eq:CollIntrinsic}\end{gather}
 Here, the first term describes spin-independent scattering, with
$K\left(\vartheta\right)=K\left(\azimuthalAngle'-\azimuthalAngle\right)=W\left(q\right)k_{\epsilon}/2\pi\hbar^{2}v_{\epsilon}$
and $q=2k_{\epsilon}\sin\left(\left|\vartheta\right|/2\right)$. The
factor $W(q)=\big\langle\left|\Vimpurity(\vec{q})\right|^{2}\big\rangle$
does not depend on the direction of the momentum transfer $\vec{q}$
since we assume that the system is macroscopically isotropic.  The
second term in Eq.~(\ref{eq:CollIntrinsic}), described by Eq.~(31)
of \refcite{ShytovKinetic}, includes two contributions, arising
from the spin-dependences of the density-of-states  and of the momentum
transfer for a fixed energy $\epsilon$. These contributions are proportional
to $K\left(\vartheta\right)$ and $\tilde{K}\left(\vartheta\right)\equiv(dK/d\vartheta)\,\tan\left(\vartheta/2\right)$,
resp., and explicitly depend on $\azimuthalAngle,\:\azimuthalAngle'$
through $\vec{b}\left(\azimuthalAngle\right)$ and $\vec{b}\left(\azimuthalAngle'\right)$.

We  consider Rashba spin-orbit interaction, and choose the $x$ axis
along the field $\vec{B}$, i.e.,\begin{align}
\vec{b}\left(\vec{k}\right) & =2\alpha\,\hat{\vec{z}}\times\vec{k}+\ZeemanX\,\hat{\vec{x}},\qquad\ZeemanX=g^{*}\mu_{\mathrm{B}}B,\label{eq:bofk}\end{align}
with Zeeman splitting $\ZeemanX$, thus $\vec{b}\left(\vec{k}\right)$
is in-plane and $\vec{E}$ and $\vec{B}$ are parallel, see Fig.~1(a).
(For $\vec{E}=E\vec{\hat{y}}$ there is $yz$ mirror symmetry and
$\polDensityElemTot_{z}$ vanishes. Thus, the $\polDensityElemTot_{z}$
term linear in $\vec{E}$ is determined only by the component $E_{x}$
parallel to $\vec{B}$.)

 Next, we write the kinetic equation~(\ref{eq:BlhsIntrinsic}) in
Fourier space by expanding the azimuthal dependence as $f\left(\azimuthalAngle\right)=\sum_{m=-\infty}^{\infty}e^{im\azimuthalAngle}\, f_{m}.$
Combining the in-plane spin distribution as $\fSpinEComponent{x}(\azimuthalAngle)+i\fSpinEComponent{y}(\azimuthalAngle)=\sum_{m}e^{im\azimuthalAngle}\,\PhiPMm{m}$,
and using the form of $\vec{M}$ given in \refcite{ShytovKinetic}
we find \cite{epapsPolarization}\begin{align}
\PhiPMdotm{m} & =i\ZeemanX\Phizm{m}+i\bRashba\Phizm{m-1}+\bRashba\Eparam\left(2+\gammaPlusTildeN\right)\delta_{m,2}\nonumber \\
 & +\ZeemanX\Eparam\left(1+\gammaHomTildeN\right)\delta_{\left|m\right|,1}-\tau^{-1}k_{m}\PhiPMm{m},\label{eq:Phipm}\\
\Phizdotm{m} & =\frac{i}{2}\,\bRashba^{*}\PhiPMm{1+m}-\frac{i}{2}\,\bRashba\PhiPMm{1-m}^{*}+i\,\frac{\ZeemanX}{2}\left(\PhiPMm{m}-\PhiPMm{-m}^{*}\right)\nonumber \\
 & +\ZeemanX\left|\bRashba\right|\tau\Eparam\,\Big(\delta_{m,0}+\frac{1}{2}\,\delta_{\left|m\right|,2}\Big)-\tau^{-1}k_{m}\Phizm{m},\label{eq:Phiz}\end{align}
 with inverse transport time $\tau^{-1}=2\pi\left(K_{0}-K_{1}\right)$
and $k_{m}=(K_{0}-K_{m})/(K_{0}-K_{1})$. Also, $\bRashba=2i\alpha k$,
so the spin-orbit field can be written as $b_{x}+ib_{y}\big|_{B=0}=\bRashba e^{i\azimuthalAngle}$.
Finally, the remaining parameters are \begin{equation}
\gammaPlusTildeN=2(k_{2}-1),\quad\;\gammaHomTildeN=\zeta+2\,\frac{\tilde{K}_{1}-\tilde{K}_{0}}{K_{0}-K_{1}}.\label{eq:GammaHomTildeVal}\end{equation}
 In the limit of small-angle scattering, $\gammaHomTildeN=\zeta+3.$
We have assumed that $\vec{B}$ is time-independent and any time-dependence
of $\vec{E}$ is slow compared to $\tau^{-1}$.

Let us consider general properties of Eqs.~(\ref{eq:Phipm})-(\ref{eq:Phiz}).
One can prove algebraically that\begin{equation}
\Phimum{z}{m}={\Phimum{z}{-m}}^{*}=\left(-1\right)^{m}\Phimum{z}{-m},\quad\PhiPM_{m}=-\left(-1\right)^{m}\PhiPM_{m}^{*}\label{eq:Symm}\end{equation}
for both the stationary regime and for transients generated by a time-dependent
electric field. {[}Arbitrary initial conditions might deviate from
Eq.~(\ref{eq:Symm}), but such deviations would decay to zero at
least as fast as the spin relaxation rate.{]} Also, these identities
directly follow from the symmetry properties of the components of
the pseudovector $\fSpinE$ for the system with the axial symmetry
$C_{\infty v}$ of the Rashba spin-orbit coupling in the fields $\vec{E},\:\vec{B}\parallel\hat{\vec{x}}$.

In particular, the symmetry of Eq.~(\ref{eq:Symm}) allows $\Phi_{0}^{z}\neq0$
in the stationary regime; therefore the spin polarization $\polDensityElemTot_{z}$
is generally finite, despite the fact that the effective field $\vec{b}(\vec{k})$
has only in-plane components. Now we can evaluate Eq.~(\ref{eq:Phipm})
for all $m$ and Eq.~(\ref{eq:Phiz}) for $m\geq0$, and eliminate
complex conjugated quantities using Eq.~(\ref{eq:Symm}). 

\emph{Isotropic scattering, parabolic bands, stationary regime.} First,
we assume isotropic scattering and parabolic bands, thus $\tilde{K}_{m}=0$
and \begin{equation}
\gammaPlusTildeN,\gammaHomTildeN=0,\qquad k_{m}=1-\delta_{m,0}.\label{eq:lblIsoCond}\end{equation}
 In this case, we solve the kinetic equations~(\ref{eq:Phipm})-(\ref{eq:Phiz})
exactly by setting $\partial\fSpinE/\partial t=0$. The stationary
solution is \begin{align}
\PhiPMm{m}^{\lblIso} & =\ZeemanX\tau\Eparam\delta_{\left|m\right|,1}+2\bRashba\tau\Eparam\left(\delta_{m,0}+\delta_{m,2}\right),\label{eq:PhipmIsoM}\\
\Phi_{m}^{z,\,\lblIso} & =0,\label{eq:PhizIsoM}\end{align}
which can be checked by inspection. The total spin polarization density
is $\polDensityTot=\polDensityEq+\polDensity$, with equilibrium contribution
$\polDensityEq\parallel\vec{B}$ and non-equilibrium contribution
$\polDensityElem_{\mu}=4\pi\int d\epsilon\:\Phimum{\mu}{0}$. Thus,
the out-of-plane polarization vanishes, $\polDensityElemTot_{z}=0$,
as one would expect from the above naive argument---even though the
symmetry allows $\polDensityElemTot_{z}\neq0$. Hence, vanishing $\polDensityElemTot_{z}$
is a property of the specific model of Eq.~(\ref{eq:lblIsoCond}).
On the other hand, even for this model, our solution is $\fSpinE(\azimuthalAngle)=4\tau\Eparam\left[\vec{b}(\azimuthalAngle)-\frac{1}{2}\ZeemanX\hat{\vec{x}}\right]\cos\azimuthalAngle$,
i.e., in addition to the total field $\vec{b}$, there is a correction
$-\frac{1}{2}\ZeemanX\hat{\vec{x}}$, indicating that spin-orbit and
external magnetic fields cannot be added. However, it  does not contribute
to the in-plane spin polarization, as it is averaged out when integrating
over $\azimuthalAngle$.

 The polarization $\polDensityElemEq_{x}$, in the absence of spin-orbit
coupling, arises from Pauli paramagnetism, $\polDensityElemEq_{x}=n_{\uparrow}-n_{\downarrow}=\frac{1}{2}\DOS(\epsilon_{\downarrow}-\epsilon_{\uparrow})=\frac{1}{2}\DOS\ZeemanX$,
with the density of states $\DOS=k_{\mathrm{F}}/\pi v_{\mathrm{F}}=m^{*}/\pi$,
Fermi momentum $k_{\mathrm{F}}$, Fermi velocity $v_{\mathrm{F}}$,
and effective mass $m^{*}$. This spin polarization does not depend
on the electric field, thus  $\Phimum{x}{0}=0$. On the other hand,
the electric field causes drift, producing an average spin-orbit splitting,
$\driftFieldy=\langle b_{y}\rangle=(1/\nTwoD)\iint d\epsilon d\azimuthalAngle\,\fDiagonalE(\epsilon,\azimuthalAngle)\, b_{y}=2\alpha eE_{x}\tau.$
By analogy to Pauli paramagnetism, one might guess that $\polDensityElem_{y}=\frac{1}{2}\DOS\langle b_{y}\rangle$.
This expectation is indeed met, because $\polDensityElem_{y}=\alpha eE_{x}\tau\DOS$
coincides with the value following from Eq.~(\ref{eq:PhipmIsoM}),
and it also agrees with known $B=0$ results \cite{VaskoPrima79,Edelstein90,ALGPspinPol91,BernevigStrain,Inoue04}.
Hence, for the model of Eq.~(\ref{eq:lblIsoCond}), the in-plane
polarization can be described in terms of the average spin-orbit field. 

In the field $\ZeemanX$, the equilibrium spin polarization per electron
is $\frac{1}{2}\DOS\ZeemanX/\nTwoD=\ZeemanX/2E_{\mathrm{F}}$, so
one expects that the drift caused by the charge current leads to a
spin current $\jSpin{x}{x}=(\jChargeComp{x}/e)\,\ZeemanX/2E_{\mathrm{F}}$.
Our results agree with this expectation; evaluating the definition
of $\jSpin{x}{x}$ by inserting $\vec{v}$  leads to $\jSpin{x}{x}=2\pi\int d\epsilon\: v_{\epsilon}\,\mathrm{Re}\left[\PhiPMm{1}+\PhiPMm{-1}+2\Eparam\ZeemanX\tau\right]$
\cite{epapsPolarization}, then Eqs.~(\ref{eq:Symm}) and~(\ref{eq:PhipmIsoM})
are used. Note that the spin current $\jSpin{x}{x}$ is well defined
for $\bRashba=0$ because spin is conserved; also our calculations
with finite $\bRashba$ result in the same $\jSpin{x}{x}$. Other
spin current components $\jSpin{\mu}{\nu}$ vanish, even for finite
$\bRashba$; for $\ZeemanX=0$ it is well-known that $\jSpin{z}{y}=0$
\cite{Inoue04,Mish04,Burkov04,RaimondiSchwab_cancellationNumerics,Chalaev2D,Dimitrova04,KrotkovDasSarma_Nonparabolicity}.

\emph{Anisotropic Scattering and Bloch equations.}  Now we consider
anisotropic scattering and/or non-parabolic bands, and also include
transients.  We consider  the {}``dirty limit,'' $\left|\bRashba\right|\ll\tau^{-1}\ll E_{\mathrm{F}}$
with constant $\vec{B}$ such that $\tauSpinz=\frac{1}{2}\tauSpinInPlane=1/\left|\bRashba\right|^{2}\tau$ 

\begin{equation}
\left|\omega\right|,\:\tau_{z}^{-1},\:\left|\ZeemanX\right|\ll\left|\bRashba\right|\ll\tau^{-1},\label{eq:BlDirtyLimitTauzConstM}\end{equation}
where $\omega$ is the characteristic frequency of the field $\vec{E}$,
and $\tauSpinz=\frac{1}{2}\tauSpinInPlane=\big(|\bRashba|^{2}\tau\big)^{-1}$
are the Dyakonov-Perel spin relaxation times. In this regime, $\Phizm{m}$
and $\PhiPMm{m}$ decay exponentially fast with increasing $|m|$,
since $\anisoPrefix\Phizm{m}/\anisoPrefix\Phizm{m-1}\sim\tau^{2}\bRashba\ZeemanX\ll1$
for $m\geq2$, and similarly for $\PhiPMm{m}$. This allows us to
solve kinetic equations~(\ref{eq:Phipm})-(\ref{eq:Phiz}) order-by-order
in the small parameter $(\tau/\tau_{z})^{1/2}$. Considering the lowest
non-vanishing order, it is sufficient to retain only equations for
$|m|\leq2$. Eliminating the $m=\pm1,\,\pm2$ components yields 
the equations of motion for $\fSpinE$ up to order $(\tau/\tau_{z})^{1/2}$~\cite{epapsPolarization}.

Finally, we evaluate the equations of motion for the total polarization
$\polDensityTot$ at low temperature $T$, taking all parameters at
the Fermi level\@. We obtain the \emph{Bloch equation} \begin{align}
\polDensityDotTot & =\left\langle \vec{b}\right\rangle \times\polDensityTot-\invTauSpinTensor\,\polDensityTot+\bGamma,\label{eq:BlVecM}\end{align}
where the spin relaxation tensor $\invTauSpinTensor$ is diagonal
with components $\{\tauSpinInPlane^{-1},\,\tauSpinInPlane^{-1},\,\tauSpinz^{-1}\}$
and $\bGamma=\big(\frac{1}{2}\nu\ZeemanX\,\tauSpinInPlane^{-1},\:\frac{1}{2}\nu\bAvgy\tauSpinInPlane^{-1},\:\frac{1}{4}\nu\ZeemanX\bAvgy\gammaHomTildeN\big)$.
Note that our proof of Eq.~(\ref{eq:BlVecM}) is valid only in linear
order in $\vec{E}$ {[}cf.\ Eq.~(\ref{eq:BlhsIntrinsic}){]}, i.e.,
products $\bAvgy\polDensityElem_{\mu}$ were disregarded.  

To develop a physical picture for this central result, we note that
Eq.~(\ref{eq:BlVecM}) is a Bloch equation, where polarization $\polDensityTot$
is generated with a rate $\Gamma$ and then precesses in the total
field $\left\langle \vec{b}\right\rangle =g^{*}\mu_{\mathrm{B}}\vec{B}+\driftField$
(Hanle effect). Most remarkably, for anisotropic scattering and/or
band nonparabolicity, the combined effect of spin-orbit and external
fields generates a spin polarization along the $z$ axis with rate
$\Gamma_{z}=\:\frac{1}{4}\nu\, g^{*}\mu_{\mathrm{B}}(\vec{B}\times\langle\vec{b}_{\mathrm{SO}}\rangle)_{z}\,\gammaHomTildeN=\frac{1}{2}\nu\alpha eE_{x}\tau\,\ZeemanX\,\gammaHomTildeN$,
i.e., \emph{perpendicular to both magnetic and spin-orbit fields}.
 This rate $\Gamma_{z}$ arises as follows. Scattering of nonequilibrium
carriers leads to an extra $k_{x}$-dependent $x$ polarization due
to the term proportional to $\gammaHomTildeN$ in Eq.~(\ref{eq:Phipm}).
On a timescale of $\tau$, this polarization then precesses around
the $y$ component of $\vec{b}_{\mathrm{SO}}$, as described by the
first two terms of Eq.~(\ref{eq:Phiz}). 

Next we consider the dc case, $\vec{\dot{\polDensityTot}}=0$. In
the lowest order in $\vec{E}$, the total spin polarization is $\polDensityElemTot_{x}=\frac{1}{2}\nu\ZeemanX$,\begin{align}
\polDensityElemTot_{y} & =\frac{1}{2}\nu\alpha\, eE_{x}\tau\left[2+\frac{\ZeemanX^{2}\tauSpinInPlane\tauSpinz}{1+\ZeemanX^{2}\tauSpinInPlane\tauSpinz}\,\gammaHomTildeN\right],\label{eq:syAnisoM}\\
\polDensityElemTot_{z} & =\frac{1}{2}\nu\alpha\, eE_{x}\tau\:\frac{\ZeemanX\tauSpinz}{1+\ZeemanX^{2}\tauSpinInPlane\tauSpinz}\,\gammaHomTildeN.\label{eq:szAnisoM}\end{align}

\noindent The first term of Eq.~(\ref{eq:syAnisoM}) arises from
Eq.~(\ref{eq:PhipmIsoM}), while the second term and $\polDensityElemTot_{z}$
are due to \emph{anisotropic scattering or nonparabolic bands}. The
dependence of $\polDensityElemTot_{z}$ on $\ZeemanX$ is in agreement
with the data in Fig.~1c of \refcite{KatoBulkPolarization04}, where
$\tauSpinGeomMean=(\tauSpinInPlane\tauSpinz)^{1/2}\approx$ 5 ns,
suggesting that our microscopic model might explain the experimental
observations.

\emph{Spin currents.} Evaluating Eq.~(\ref{eq:Phipm}) for $m=0$,
we find that $\Phi_{-1}^{z}=-(\ZeemanX/\bRashba)\Phi_{0}^{z}$. We
then evaluate the spin current at $T=0$, $\jSpin{z}{y}=4\pi\int d\epsilon\: v_{\epsilon}\,\mathrm{Im}\Phimum{z}{-1}=4\pi\int d\epsilon\: v_{\epsilon}(\ZeemanX/|\bRashba|)\,\Phi_{0}^{z}=v_{\mathrm{F}}(\ZeemanX/|\bRashba|)\, s_{z}$,
finding that $\jSpin{z}{y}$ is proportional to $s_{z}$. (This relationship
also follows from the Heisenberg equation of $\dot{\sigma}_{y}$.)
Hence, the polarization $s_{z}$ {[}Eq.~(\ref{eq:szAnisoM}){]} leads
to a transverse spin current $\jSpin{z}{y}$; a finite $\jSpin{z}{y}$
is in agreement with numerical results of \refcite{LinLuiLei06}.
For $\ZeemanX=0$, this relation is equivalent to the argument \cite{Dimitrova04,Chalaev2D}
based on equations of motion \cite{Burkov04}, showing that $j_{y}^{z}=0$.

\emph{Spin Dynamics}. Even for isotropic scattering, a time-dependent
electric field leads to an out-of-plane polarization $\polDensityElemTot_{z}(\omega)=\frac{i}{2}\omega\DOS\ZeemanX\driftFieldy(\omega)/[\ZeemanX^{2}-\omega^{2}+\tauSpinGeomMean^{-2}-i(\tauSpinInPlane^{-1}+\tauSpinz^{-1})\omega]$;
however, it has no static component $\polDensityElemTot_{z}(\omega=0)$.
Similar results were found in \refcite{DLesr} for $\left|\bRashba\right|\ll\ZeemanX,\,\tau^{-1}$.

Spin dynamics is accessible in a pump-probe scheme \cite{KatoBulkPolarization04}.
Namely, spins can be pumped by applying a short electric pulse of
duration $t_{\mathrm{p}}\ll\tauSpinz,\,\ZeemanX^{-1}$. Then, according
to Eq.~(\ref{eq:BlVecM}), the spin polarization immediately after
the pulse is $s_{z}(0)=t_{\mathrm{p}}\,\Gamma_{z}\propto\ZeemanX\gammaHomTildeN$,
i.e., $s_{z}(0)$ is an odd function of $\ZeemanX$. Solving the Bloch
equation (\ref{eq:BlVecM}), we get\begin{equation}
\polDensityElemTot_{z}(t)=\polDensityElemTot_{z}(0)\, e^{-3t/4\tauSpinz}\left[\cos\Omega t-\frac{1+2/\gammaHomTildeN}{4\Omega\tauSpinz}\,\sin\Omega t\right]\end{equation}
 with frequency $\Omega=\sqrt{(4\ZeemanX\tauSpinz)^{2}-1}/4\tauSpinz$
of Hanle oscillations (for consistency, we only consider terms linear
in $\vec{E}$). We plot $\polDensityElemTot_{z}(t)$ in Fig.~1(b),
taking the parameters of \refcite{KatoBulkPolarization04} and with
a choice of $\gammaHomTildeN=0.03$, and find qualitative agreement
with the experiment. The weak-field region $4|\ZeemanX|\tau_{z}<1$,
where the oscillations are overdamped, is very narrow, $|B|\lesssim0.25\:\mathrm{mT}$.
Note that the experimental data shows that the sign of $\polDensityElemTot_{z}$
depends on the sign of $\ZeemanX$, already on time scales much shorter
than $|\ZeemanX|^{-1}$. Therefore, the sign of $\polDensityElemTot_{z}$
cannot be due to spin precession in the external magnetic field---implying
that a polarization generation mechanism like the one described above
was experimentally observed in \refcite{KatoBulkPolarization04}.

Strictly speaking, quantitative comparison with the data of \refcite{KatoBulkPolarization04}
cannot be performed because the films were of low mobility $E_{\mathrm{F}}\tau\sim1$,
violating the assumptions of our Boltzmann description, and were in
3D regime (a coupling $k_{y}\sigma_{x}-k_{x}\sigma_{y}$ occurs here
due to strain). Furthermore, in models with a more complicated spin-orbit
interaction than the Rashba coupling, other sources of $z-$polarization
might become important. However, Eq.~(\ref{eq:BlDirtyLimitTauzConstM})
was satisfied, because $\hbar/\tau_{z}\sim3\times10^{-8}\:\mathrm{eV}$;
$\left|\ZeemanX\right|\lesssim10^{-6}\:\mathrm{eV}$; $|b_{\alpha}|\sim10^{-5}\:\mathrm{eV}$;
and $\hbar/\tau\sim2\times10^{-3}$ eV \cite{KatoBulkPolarization04,KatoCSpin04}.

The effective field $\vec{b}(\vec{k})$ for a 2DEG with pure linear
Dresselhaus coupling, on the $(001)$ surface of a III-V material,
is obtained by replacing $\vec{k}$ on the right hand side of Eq.~(\ref{eq:bofk})
by $\vec{q}\equiv\mathcal{R}\vec{k}$, where $\mathcal{R}$ denotes
reflection through the $(110)$ crystal plane. Our result~(\ref{eq:szAnisoM})
for the polarization $s_{z}$ can be applied to this case if we replace
$E_{x}$ by the component of the electric field along the direction
$\vec{B}'\equiv\mathcal{R}\vec{B}$. For general forms of the spin
orbit coupling, we note that the $\mathrm{C}_{2v}$ symmetry of the
system ensures that if $B=0$, there can be no term in $s_{z}$ linear
in $\vec{E}$. However, there could be terms non-linear in $\vec{E}$,
if $\vec{E}$ is not parallel to a symmetry direction $[110]$ or
$[1\bar{1}0]$, e.g. $s_{z}\propto E_{x}^{2}-E_{y}^{2}$ where $x$
refers to the $[100]$ crystal axis, which would then give an all-electrical
mechanism for generating out of plane spin polarization. 

In conclusion, we proposed a mechanism for generating bulk spin populations
polarized perpendicularly to magnetic and spin-orbit fields; for 2D
systems this is an out-of-plane polarization. It relies on anisotropic
impurity scattering and/or band nonparabolicity and provides a new
method for electrical control of electron spins. Our model is derived
for 2D systems, but the results should have a more general validity,
and they agree with recent observations of combined effects of the
external magnetic and spin-orbit fields in 3D samples.

We thank A.H. MacDonald for attracting our attention to the intriguing
results of \refcite{KatoBulkPolarization04} and acknowledge discussions
with A.A. Burkov,  D. Loss, and B. Rosenow. This work was supported
by NSF Grants No. DMR-02-33773, No. PHY-01-17795, and No. PHY99-07949,
and by the Harvard Center for Nanoscale Systems.

\bibliographystyle{myapsrev}
\bibliography{S:/Physics/references}

\begin{thebibliography}{26}
\bibitem[{foo()}]{footnoteAnisobdr}
\bibinfo{note}{For anisotropic scattering, defining $\driftField$ is less
  straightforward \cite{ALGPspinPol91}.}
\end{thebibliography}
\clearpage

\newpage
{}{}

\newpage
\setcounter{page}{1}

\setcounter{equation}{0}

\renewcommand{\theequation}{S\arabic{equation}}

\textbf{\large Supplemental Material}{\large \par}

\newcommand{\PhiPMmtau}[2]{\PhiPM_{#1}^{(#2)}}

\newcommand{\PhiPMdotmtau}[2]{\PhiPMdotm{#1}^{(#2)}}

\newcommand{\Phixdotmtau}[2]{\dot{\Phi}_{#1}^{x(#2)}}

\newcommand{\Phixmtau}[2]{\Phi_{#1}^{x(#2)}}

\newcommand{\Phizmtau}[2]{\Phi_{#1}^{z(#2)}}

\newcommand{\Phizdotmtau}[2]{\dot{\Phi}_{#1}^{z(#2)}}

\newcommand{\Phiymtau}[2]{\Phi_{#1}^{y(#2)}}

\newcommand{\Phiydotmtau}[2]{\dot{\Phi}_{#1}^{y(#2)}}

\newcommand{\FTarrowDisplayed}{\stackrel{\mathrm{FT}}{\longrightarrow}}

\newcommand{\EparamPlusAIscattering}{\tilde{\Eparam}_{\left(+\right)}}

\newcommand{\EparamPlusBextScattering}{\tilde{\Eparam}_{0}}

\newcommand{\EparamSum}{\beta_{\Sigma}}

\newcommand{\bFTComp}[2]{b_{#2}^{#1}}

\newcommand{\jSpLblPhimu}{(1)}

\newcommand{\jSpLblPhiC}{(2)}

\section{List of symbols}

\begin{longtable}{rp{71mm}}
$\energyOfK{k}$&
Dispersion law in absence of spin-orbit interaction\tabularnewline
$\bsigma$&
Vector of Pauli matrices, $(\sigma_{x},\,\sigma_{y},\,\sigma_{z})$\tabularnewline
$\vec{b}\left(\vec{k}\right)$&
Total field in energy units, containing both spin-orbit and external
magnetic fields\tabularnewline
$\vec{E}$&
Electric field, $\vec{E}=E\hat{\vec{x}}$\tabularnewline
$\vec{B}$&
External magnetic field, $\vec{B}=B\hat{\vec{x}}$\tabularnewline
$\ZeemanX$&
Zeeman splitting, $\ZeemanX=g^{*}\mu_{\mathrm{B}}B$, with effective
$g$-factor $g*$ and Bohr magneton $\mu_{\mathrm{B}}$\tabularnewline
$e$&
charge of carrier, for electrons $e<0$\tabularnewline
$\alpha$&
Rashba spin-orbit coupling constant, $\bRashba=2i\alpha k$\tabularnewline
$V_{\mathrm{i}}\left(\vec{r}\right)$&
Impurity potential\tabularnewline
$\vec{v}$&
Spin-dependent velocity, $\vec{v}=i\left[H,\,\vec{r}\right]$\tabularnewline
$\left\{ \,,\,\right\} $&
Anticommutator, $\left\{ A,\, B\right\} =AB+BA$\tabularnewline
$\polDensityTot$&
Spin polarization density, $\polDensityTot\left(\vec{r}\right)=\nTwoD\left\langle \bsigma\right\rangle =\polDensityEq+\polDensity$,
containing both equilirum and non-equilibrium contributions $\polDensityEq$
and $\polDensity$, resp.\tabularnewline
$\vec{j}^{\mu}$&
Spin current, $\vec{j}^{\mu}\left(\vec{r}\right)=\nTwoD\langle{\textstyle \frac{1}{2}}\,\{\sigma_{\mu},\:\vec{v}\}\rangle$.\tabularnewline
$\vec{k}$&
Wave vector. In two dimensions, $\vec{k}=(k_{x},\: k_{y})=k\kUnitVec=(k\cos\azimuthalAngle,\: k\sin\azimuthalAngle)$ \tabularnewline
$\hat{f}(\vec{k})$&
Distribution function as $2\times2$ matrix, $\hat{f}(\vec{k})=\hat{f}_{0}(\vec{k})+\frac{1}{2}\fDiagonal(\vec{k})\,1\!\!1+\fSpin(\vec{k})\cdot\bsigma$,
as a function of wave vector $\vec{k}$\tabularnewline
$\hat{f}_{0}(\vec{k})$&
Equilibrium distribution function, spin-dependent due to magnetic
field\tabularnewline
$\fDiagonal(\vec{k})$&
Non-equilibrium particle density\tabularnewline
$\fSpin(\vec{k})$&
Non-equilibrium spin polarization density\tabularnewline
$k_{\epsilon}$&
Spin-independent wave number contribution for given energy $\epsilon$,
i.e., $\energyOfK{k_{\epsilon}}=\epsilon$\tabularnewline
$v_{\epsilon}$&
Spin-independent velocity contribution, $v_{\epsilon}=\energyOfK{k_{\epsilon}}'/\hbar$\tabularnewline
$\zeta$&
Band non-parabolicity, $1+\zeta=(k_{\epsilon}/v_{\epsilon})(\partial v_{\epsilon}/\partial k_{\epsilon})$\tabularnewline
$\DOS$&
Density of states at Fermi level\tabularnewline
$\fMatrixE\left(\azimuthalAngle,\,\epsilon\right)$&
Non-equilibrium distribution function $\fMatrixE=\fMatrixE\left(\azimuthalAngle,\,\epsilon\right)$
as $2\times2$ matrix, as a function of direction of $\vec{k}$ and
energy $\epsilon$, $\fMatrixE=\frac{1}{2}\fDiagonalE\,1\!\!1+\fSpinE\cdot\bsigma$\tabularnewline
$\fDiagonalE\left(\azimuthalAngle,\,\epsilon\right)$&
Excess particle density, as a function of $\azimuthalAngle$ and $\epsilon$,
$\fDiagonalE=8\Eparam\tau v_{\epsilon}k_{\epsilon}\cos\azimuthalAngle$\tabularnewline
$\fSpinE\left(\azimuthalAngle,\,\epsilon\right)$&
Non-equilibrium spin polarization density, as a function of $\azimuthalAngle$
and $\epsilon$\tabularnewline
$K\left(\vartheta\right)$&
Angular dependence of spin-independent scattering, in Born approximation
$K\left(\azimuthalAngle'-\azimuthalAngle\right)=W\left(q\right)k_{\epsilon}/2\pi\hbar^{2}v_{\epsilon}$\tabularnewline
$\vartheta$&
Scattering angle, $\vartheta=\azimuthalAngle'-\azimuthalAngle$\tabularnewline
$\vec{q}$&
Momentum transfer, $q=2k_{\epsilon}\sin\left(\left|\vartheta\right|/2\right)$. \tabularnewline
$\tilde{K}\left(\vartheta\right)$&
Scattering contribution due to spin-dependence of momentum transfer,
$\tilde{K}\left(\vartheta\right)=(dK/d\vartheta)\,\tan\left(\vartheta/2\right)$\tabularnewline
$\Phizm{m}$&
Fourier coefficients of $\fSpinEComponent{z}(\azimuthalAngle)$\tabularnewline
$\PhiPMm{m}$&
Fourier coefficients of $\fSpinEComponent{x}(\azimuthalAngle)+i\fSpinEComponent{y}(\azimuthalAngle)$\tabularnewline
$K_{m},\,\tilde{K}_{m}$&
Fourier coefficients of $K\left(\vartheta\right)$ and $\tilde{K}\left(\vartheta\right)$,
resp.\tabularnewline
$k_{m}$&
Defined as $k_{m}=(K_{0}-K_{m})/(K_{0}-K_{1})$, i.e., $k_{m}>0$
for $m>0$\tabularnewline
$\tau$&
Transport lifetime, $\tau^{-1}=2\pi(K_{0}-K_{1})$\tabularnewline
$\Eparam$&
Factor describing coupling to electric field, $\Eparam=(eE/16\pi^{2}v_{\epsilon})\,(-\partial f_{0}/\partial\epsilon)$\tabularnewline
$f_{0}(\epsilon)$&
Fermi distribution function\tabularnewline
$\gammaPlusTildeN$&
Spin-dependent collision contribution due to $\bRashba$, $\gammaPlusTildeN=2(k_{2}-1)$\tabularnewline
$\gammaHomTildeN$&
Spin-dependent collision contribution due to $\vec{B}$, $\gammaHomTildeN=\zeta+2\,(\tilde{K}_{1}-\tilde{K}_{0})/K_{0}-K_{1}$\tabularnewline
$c$&
Coupling coefficient in kinetic equations,  $c=i\bRashba^{*}\ZeemanX\tau\Eparam$\tabularnewline
$\tau_{\mathrm{z}}$&
Dyakonov-Perel spin relaxation time, $\tau_{\mathrm{z}}=1/\left|\bRashba\right|^{2}\tau$\tabularnewline
$\tauSpinInPlane$&
In-plane spin relaxation time, $\tauSpinInPlane=2\tauSpinz$\tabularnewline
$\invTauSpinTensor$&
Spin relaxation tensor \tabularnewline
$\bGamma$&
Spin generation with rate $\Gamma$\tabularnewline
\end{longtable}

\section{Kinetic equation in Fourier space}

In Fourier space, the kinetic equation Eq.~(\ref{eq:BlhsIntrinsic})
becomes, for $m=0$, (note that $k_{0}=0$)\begin{align}
\PhiPMdotm{0}= & i\ZeemanX\Phizm{0}+i\bRashba\Phizm{-1},\label{eq:Phip0}\\
\Phizdotm{0}= & \frac{i}{2}\,\bRashba^{*}\PhiPMm{1}-\frac{i}{2}\,\bRashba\PhiPMm{1}^{*}+c-\frac{\ZeemanX}{2i}\left(\PhiPMm{0}-\PhiPMm{0}^{*}\right),\label{eq:Phiz0}\end{align}
while for $m\neq0$:\begin{align}
\PhiPMdotm{m}+\tau^{-1}k_{m}\PhiPM_{m} & =i\ZeemanX\Phizm{m}+i\bRashba\Phizm{m-1}+2\bRashba\EparamPlusAIscattering\,\delta_{m,2}\nonumber \\
 & \qquad+\ZeemanX\EparamPlusBextScattering\,\delta_{\left|m\right|,1},\label{eq:Phipmm}\\
\Phizdotm{m}+\tau^{-1}k_{m}\Phizm{m} & =\frac{i}{2}\,\bRashba^{*}\PhiPM_{1+m}-\frac{i}{2}\,\bRashba\PhiPM_{1-m}^{*}-\frac{\ZeemanX}{2i}\nonumber \\
 & \qquad\times\left(\PhiPM_{m}-\PhiPM_{-m}^{*}\right)+\frac{c}{2}\,\delta_{\left|m\right|,2},\label{eq:Phizm}\end{align}
with $\EparamPlusAIscattering=\Eparam+\frac{1}{2}\Eparam\gammaPlusTildeN=k_{2}\beta$
and with $\EparamPlusBextScattering=\Eparam+\Eparam\gammaHomTildeN.$

The contributions proportional to $\gammaPlusTildeN$ and $\gammaHomTildeN$
arise from the second term in Eq.~(\ref{eq:CollIntrinsic}), where
the kernel $\vec{M}$ is adopted from \refcite{ShytovKinetic}, \begin{align}
\vec{M}\!\left(\azimuthalAngle,\azimuthalAngle'\right) & =\frac{v_{\epsilon}}{4k_{\epsilon}}\, K\!\left(\vartheta\right)\frac{\partial}{\partial\epsilon}\left[\frac{k_{\epsilon}\vec{b}\!\left(\azimuthalAngle\right)}{v_{\epsilon}}\right]+\frac{\vec{b}\!\left(\azimuthalAngle\right)+\vec{b}\!\left(\azimuthalAngle'\right)}{4\hbar k_{\epsilon}v_{\epsilon}}\,\tilde{K}\!\left(\vartheta\right).\end{align}
Concretely, we find $\gammaPlusTildeN=2\pi\tau(\tilde{K}_{2}-\tilde{K}_{0})$
and $\gammaHomTildeN=\zeta+4\pi\tau(\tilde{K}_{1}-\tilde{K}_{0})$.%
\footnote{In terms of notation used in \refcite{ShytovKinetic}, we see that
$\gammaPlusTildeN=\gamma_{\left(+\right)}/\eta_{0}$ for winding numer
$N=1$, corresponding to the field $\bRashba$; whereas $\gammaHomTildeN=\gamma_{\left(+\right)}/\eta_{0}\,\big|_{N=\tilde{N}=0}$
for the field $B$.%
} By explicit evaluation of $\tilde{K}_{m}$, we find the relations
$\tilde{K}_{0}=2\sum_{n>0}(-1)^{n}nK_{n}$ and $\tilde{K}_{m}=\left(-1\right)^{m}\tilde{K}_{0}-\left|m\right|K_{m}-2\sum_{0<n<\left|m\right|}\left(-1\right)^{m+n}nK_{n}$,
in particular, $\tilde{K}_{0}-\tilde{K}_{1}=2\tilde{K}_{0}+K_{1}$
and $\tilde{K}_{2}-\tilde{K}_{0}=2(K_{1}-K_{2})$, see Sec.~\ref{sub:tildeKtoK}.
This allows us to transform $\gammaPlusTildeN$ and $\gammaHomTildeN$
and we obtain Eq.~(\ref{eq:GammaHomTildeVal}).

\subsection{Solution for isotropic scattering}

We consider the stationary case $\partial\Phi_{m}/\partial t=0$,
isotropic scattering, and parabolic bands, thus $\gammaPlusTildeN,\gammaHomTildeN=0$
and $\EparamPlusBextScattering=\Eparam$. We find the solution of
the kinetic equation\begin{align}
\PhiPMm{m} & =\ZeemanX\tau\Eparam\delta_{\left|m\right|,1}+2\bRashba\tau\Eparam\left(\delta_{m,0}+\delta_{m,2}\right),\label{eq:PhipmIso}\\
\Phizm{m} & =0.\label{eq:PhizIso}\end{align}

We now prove that Eqs.~(\ref{eq:PhipmIso})-(\ref{eq:PhizIso}) satisfy
Eqs.~(\ref{eq:Phip0})-(\ref{eq:Phizm}). Eq.~(\ref{eq:Phip0})
is trivially satisfied. Next, we use $c=-i\bRashba\ZeemanX\tau\Eparam$
and write \begin{equation}
\PhiPMm{m}=\frac{ic}{\bRashba}\delta_{\left|m\right|,1}+\frac{2ic}{\ZeemanX}\left(\delta_{m,0}+\delta_{m,2}\right),\end{equation}
 and insert into the l.h.s. of Eq.~(\ref{eq:Phiz0}), \begin{align}
 & \frac{i}{2}\,\bRashba^{*}\frac{ic}{\bRashba}-\frac{i}{2}\,\bRashba\frac{ic}{\bRashba}+c-\frac{\ZeemanX}{2i}\left(\frac{2ic}{\ZeemanX}+\frac{2ic}{\ZeemanX}\right)\\
 & =\left(\frac{1}{2}+\frac{1}{2}+1-1-1\right)c=0.\end{align}
For $m\neq0$, we insert $\Phizm{m}=0$ in Eq.~(\ref{eq:Phipmm}),
use $\EparamPlusAIscattering=k_{2}\beta$, and obtain \begin{equation}
\tau^{-1}k_{m}\PhiPMm{m}=2\bRashba k_{2}\Eparam\,\delta_{m,2}+\ZeemanX\Eparam\,\delta_{\left|m\right|,1},\end{equation}
which by comparison with Eq.~(\ref{eq:PhipmIso}) is also satisfied
(because $k_{1}=1$ by definition). Finally, evaluating Eq.~(\ref{eq:Phizm})
and dividing by $c$, we obtain \begin{align}
0 & =-\frac{i}{2}\,\left(i\delta_{\left|m+1\right|,1}+\bRashba\frac{2i}{\ZeemanX}\delta_{\left|m\right|,1}\right)\nonumber \\
 & \qquad-\frac{i}{2}\,\left(i\delta_{\left|m-1\right|,1}-\bRashba\frac{2i}{\ZeemanX}\delta_{\left|m\right|,1}\right)\nonumber \\
 & \qquad-\frac{\ZeemanX}{2i}\left(\frac{2i}{\ZeemanX}\delta_{m,2}+\frac{2i}{\ZeemanX}\delta_{-m,2}\right)+\frac{1}{2}\,\delta_{\left|m\right|,2}\\
 & =\frac{1}{2}\delta_{\left|m+1\right|,1}+\frac{1}{2}\delta_{\left|m-1\right|,1}-\delta_{m,2}-\delta_{m,-2}+\frac{1}{2}\delta_{\left|m\right|,2}\\
 & \stackrel{m\neq0}{=}\frac{1}{2}\delta_{m,-2}+\frac{1}{2}\delta_{m,2}-\delta_{m,2}-\delta_{m,-2}+\frac{1}{2}\delta_{\left|m\right|,2}=0;\end{align}
therefore, our solution for $\Phi$ is valid. Note that the property
$k_{m}=1-\delta_{m,0}$ for isotropic scattering was not used explicitly
in the above proof. 

Equations~(\ref{eq:PhipmIso})-(\ref{eq:PhizIso}) correspond to
\begin{align}
\Phi_{x}\left(\azimuthalAngle\right) & =2\tau\Eparam\,\left(\ZeemanX-2\left|\bRashba\right|\sin\azimuthalAngle\right)\cos\azimuthalAngle=4\tau\Eparam\: b_{\Sigma}^{x}\left(\azimuthalAngle\right)\cos\azimuthalAngle,\label{eq:PhixThetaIso}\\
\Phi_{y}\left(\azimuthalAngle\right) & =4\tau\left|\bRashba\right|\Eparam\:\cos^{2}\azimuthalAngle=4\tau\Eparam\: b_{\Sigma}^{y}\left(\azimuthalAngle\right)\cos\azimuthalAngle,\\
\Phi_{z}\left(\azimuthalAngle\right) & =0.\end{align}
As for the non-equilibrium charge distribution $\fDiagonalE$, a factor
of $\cos\azimuthalAngle$ is present. Remarkably, $\vec{b}_{\Sigma}\left(\azimuthalAngle\right)=\vec{b}\left(\azimuthalAngle\right)+\frac{1}{2}\, g^{*}\mu_{\mathrm{B}}\vec{B}$
differs from the total field $\vec{b}$ {[}Eq.~(\ref{eq:bofk}){]}
that enters in the Hamiltonian.

\section{Effective Bloch equations }

In the following, we consider the dirty regime and solve the kinetic
equations in orders of the small parameter $(\tau/\tauSpinz)^{1/2}$.
The contributions to $\PhiPMm{m}$ and $\Phizm{m}$ of order $\tau^{n}$
are denoted as $\PhiPMmtau{m}{n}$ and $\Phizmtau{m}{n}$, resp. It
is convenient to choose units such that $\tauSpinz$ is of order unity.
Let us now consider the regime, \begin{equation}
\left|\omega\right|,\:\tau_{z}^{-1},\:\left|\ZeemanX\right|\ll\left|\bRashba\right|\ll\tau^{-1},\label{eq:BlDirtyLimitTauzConst}\end{equation}
i.e., in Fourier space with respect to $t$, $\PhiPMdotmtau{m}{n}\to-i\omega\PhiPMmtau{m}{n}(\omega)$
is of the same order in $\tau$ as $\PhiPMmtau{m}{n}$.%
\footnote{The following derivation becomes simpler if we first insert the ansatz
$\fSpinE=\fSpinE^{\lblIso}+\anisoPrefix\fSpinE$. This replaces $\PhiPMm{m}\to\anisoPrefix\PhiPMm{m}$;
$\Phizm{m}\to\anisoPrefix\Phizm{m}$; $\EparamPlusAIscattering\to0$;
and $\EparamPlusBextScattering\to\Eparam\gammaHomTildeN$ in the following
equations and $c\to0$ in Eq.~(\ref{eq:Phiz12Raw}).%
}

We take order $O(\tau^{-1/2})$ of Eq.~(\ref{eq:Phipmm}) for $m=2$\begin{align}
\PhiPMmtau{2}{1/2} & =2\bRashba\tau k_{2}^{-1}\EparamPlusAIscattering,\label{eq:BlPhiPMm12}\end{align}
and order $O(\tau^{0})$ of Eq.~(\ref{eq:Phizm}) for $m=1$,\begin{align}
\Phizmtau{1}{1} & =\frac{i}{2}\bRashba^{*}\tau\PhiPMmtau{2}{1/2}-\frac{i}{2}\bRashba\tau{\PhiPMmtau{0}{1/2}}^{*}\\
 & \stackrel{(\ref{eq:BlPhiPMm12})}{=}-\frac{i}{2}\bRashba\tau{\PhiPMmtau{0}{1/2}}^{*}+i\left|\bRashba\right|^{2}\tau^{2}k_{2}^{-1}\EparamPlusAIscattering,\label{eq:BlPhizm1tau1}\end{align}
where we have indicated above the equality sign that we used Eq.~(\ref{eq:BlPhiPMm12}).
Taking Eq.~(\ref{eq:Phip0}) for order $O(\tau^{1/2})$ and using
that $\Phizm{-1}=\left(\Phizm{1}\right)^{*}$ yields\begin{align}
\PhiPMdotmtau{0}{1/2} & \stackrel{(\ref{eq:Phip0})}{=}i\ZeemanX\Phizmtau{0}{1/2}+i\bRashba\Phizmtau{-1}{1}\\
 & \stackrel{(\ref{eq:BlPhizm1tau1})}{=}i\ZeemanX\Phizmtau{0}{1/2}-\frac{1}{2}\left|\bRashba\right|^{2}\tau\,\PhiPMmtau{0}{1/2}\nonumber \\
 & \qquad+\bRashba\left|\bRashba\right|^{2}\tau^{2}k_{2}^{-1}\EparamPlusAIscattering\\
 & =i\ZeemanX\Phizmtau{0}{1/2}-\tauSpinInPlane^{-1}\,\PhiPMmtau{0}{1/2}+2\tauSpinInPlane^{-1}\bRashba\tau\Eparam,\label{eq:PhiPMdotDirt}\end{align}
where we used $\tauSpinInPlane^{-1}=\frac{1}{2}\tauSpinz^{-1}=\frac{1}{2}\left|\bRashba\right|^{2}\tau$
and $k_{2}^{-1}\EparamPlusAIscattering=\Eparam$.

Next, we derive the equation of motion for $\Phizdotm{0}$. To this
end, we take order $O(\tau^{0})$ of Eq.~(\ref{eq:Phipmm}) for $m=1$,
\begin{align}
\PhiPMmtau{1}{1} & =i\bRashba\tau\Phizmtau{0}{1/2}+\ZeemanX\tau\EparamPlusBextScattering.\label{eq:BlPhiPMm1Ord1}\end{align}
Then we take order $O(\tau^{1/2})$ of Eq.~(\ref{eq:Phiz0}), \begin{align}
\Phizdotmtau{0}{1/2} & =\frac{i}{2}\,\bRashba^{*}\PhiPMmtau{1}{1}-\frac{i}{2}\,\bRashba{\PhiPMmtau{1}{1}}^{*}+c\nonumber \\
 & \qquad+\frac{i}{2}\ZeemanX\left(\PhiPMmtau{0}{1/2}-{\PhiPMmtau{0}{1/2}}^{*}\right)\label{eq:Phiz12Raw}\\
 & \stackrel{(\ref{eq:BlPhiPMm1Ord1})}{=}-\left|\bRashba\right|^{2}\tau\Phizmtau{0}{1/2}+\frac{i}{2}\,\ZeemanX\tau\left(\bRashba^{*}-\bRashba\right)\EparamPlusBextScattering+c\nonumber \\
 & \qquad-\ZeemanX\:\mathrm{Im}\PhiPMmtau{0}{1/2}\\
 & =-\left|\bRashba\right|^{2}\tau\Phizmtau{0}{1/2}-\ZeemanX\:\mathrm{Im}\PhiPMmtau{0}{1/2}+c\left(2+\gammaHomTildeN\right),\label{eq:PhizdotDirt}\end{align}
where we used $\frac{i}{2}\,\ZeemanX\tau\left(\bRashba^{*}-\bRashba\right)=i\bRashba^{*}\ZeemanX\tau=c/\Eparam$.

Finally, using $i\bRashba^{*}\left|\bRashba\right|^{2}\tau^{2}\Eparam=2i\bRashba^{*}\tauSpinInPlane^{-1}\tau\Eparam=2c\tauSpinInPlane^{-1}/\ZeemanX$,
we obtain the \emph{Bloch equations} for the nonequilibrium distribution
function $\fSpinE$,\begin{align}
\Phiydotmtau{0}{1/2} & \stackrel{(\ref{eq:PhiPMdotDirt})}{=}\ZeemanX\Phizmtau{0}{1/2}-\tauSpinInPlane^{-1}\,\Phiymtau{0}{1/2}+2c\tauSpinInPlane^{-1}/\ZeemanX,\label{eq:BlPhiy}\\
\Phizdotmtau{0}{1/2} & \stackrel{(\ref{eq:PhizdotDirt})}{=}-\ZeemanX\Phiymtau{0}{1/2}-\tauSpinz^{-1}\Phizmtau{0}{1/2}+c\left(2+\gammaHomTildeN\right).\label{eq:BlPhiz}\end{align}
Therefore, the spin-orbit field generates $y$ polarization with rate
$\tauSpinz^{-1}\left|\bRashba\right|\tau\Eparam$, whereas the combined
effect of $\vec{b}_{\mathrm{SO}}$ and $\vec{B}$ generates $z$ polarization
with rate $c(2+\gammaHomTildeN)$, which is proportional to $\ZeemanX$.

\subsection{Bloch equations for total polarization $\polDensityTot$}

To obtain the spin polarzation $\polDensityTot$, we integrate the
equations for the total non-equilibrium contribution $\fSpinE$ at
temperature $T=0$, see Sec.~\ref{ssec:polarizationFT}. Noting that
the drift field is $\bAvgy=2\alpha eE\tau$, we find \begin{align}
4\pi\int d\epsilon\:2c & =16\pi\int d\epsilon\alpha k_{\epsilon}\ZeemanX\tau\Eparam=\frac{k_{\mathrm{F}}}{\pi v_{\mathrm{F}}}\alpha\ZeemanX eE\tau\\
 & =\nu\ZeemanX\alpha eE\tau=\bAvgy\,\frac{1}{2}\nu\ZeemanX=\bAvgy\,\polDensityElemEq_{x}.\end{align}
Therefore, for the spin polarization $\polDensityElem_{\mu}\stackrel{(\ref{eq:smuToFT})}{=}4\pi\int d\epsilon\:\Phimum{\mu}{0}$,
we obtain \begin{align}
\polDensityElemDot_{x} & \stackrel{(\ref{eq:PhiPMdotDirt})}{=}-\tauSpinInPlane^{-1}\,\polDensityElem_{x},\label{eq:Blsx}\\
\polDensityElemDot_{y} & \stackrel{(\ref{eq:BlPhiy})}{=}\ZeemanX\,\polDensityElem_{z}-\tauSpinInPlane^{-1}\,\polDensityElem_{y}+\frac{1}{2}\nu\bAvgy\tauSpinInPlane^{-1},\label{eq:Blsy}\\
\polDensityElemDot_{z} & \stackrel{(\ref{eq:BlPhiz})}{=}\bAvgy\,\polDensityElemEq_{x}-\ZeemanX\,\polDensityElem_{y}-\tauSpinz^{-1}\,\polDensityElem_{z}+\frac{1}{4}\nu\bAvgy\ZeemanX\,\gammaHomTildeN,\label{eq:Blsz}\end{align}
where we have used $\frac{eE\tau}{4\pi v_{\mathrm{F}}}\, i\bRashba^{*}\ZeemanX\,\gammaHomTildeN=\frac{k_{\mathrm{F}}}{4\pi v_{\mathrm{F}}}2\alpha eE\tau\:\ZeemanX\,\gammaHomTildeN=\frac{1}{4}\nu\bAvgy\ZeemanX\,\gammaHomTildeN$.
The stationary solution is \begin{align}
\polDensityElem_{y} & =\frac{1}{4}\nu\bAvgy\left[2+\frac{\ZeemanX^{2}\tauSpinInPlane\tauSpinz}{1+\ZeemanX^{2}\tauSpinInPlane\tauSpinz}\,\gammaHomTildeN\right],\label{eq:syAniso}\\
\polDensityElem_{z} & =\frac{1}{4}\nu\bAvgy\:\frac{\ZeemanX\tauSpinz}{1+\ZeemanX^{2}\tauSpinInPlane\tauSpinz}\,\gammaHomTildeN.\label{eq:szAniso}\end{align}

We assume that $\ZeemanX$ does not change over time, thus $\partial\polDensityElemEq/\partial t=0$.
The equilibrium polarization is $\polDensityElemEq_{y}=\polDensityElemEq_{z}=0$
and $\polDensityElemEq_{x}=\frac{1}{2}\nu\ZeemanX$, thus $0=-\tauSpinInPlane^{-1}\polDensityElemEq_{x}+\frac{1}{2}\nu\ZeemanX\,\tauSpinInPlane^{-1}$,
which can then be added to the r.h.s.\ of Eq.~(\ref{eq:Blsx}).
This leads to the Bloch equations for the total spin polarization
$\polDensityElemTot=\polDensityElem+\polDensityElemEq$, \begin{align}
\polDensityElemDotTot_{x} & =-\tauSpinInPlane^{-1}\,\polDensityElemTot_{x}+\frac{1}{2}\nu\ZeemanX\,\tauSpinInPlane^{-1},\label{eq:BlsxTot}\\
\polDensityElemDotTot_{y} & =\ZeemanX\,\polDensityElemTot{}_{z}-\tauSpinInPlane^{-1}\,\polDensityElemTot_{y}+\frac{1}{2}\nu\bAvgy\tauSpinInPlane^{-1},\label{eq:BlsyTot}\\
\polDensityElemDotTot_{z} & =\left(\polDensityElemTot_{x}-\polDensityElem_{x}\right)\,\bAvgy-\ZeemanX\,\polDensityElemTot_{y}-\tauSpinz^{-1}\,\polDensityElemTot_{z}+\frac{1}{4}\nu\bAvgy\ZeemanX\,\gammaHomTildeN.\label{eq:BlszTot}\end{align}
In linear order in $E$, $\bAvgy\polDensityElem_{\mu}$ vanishes and
we can add the terms $-\bAvgy\polDensityElem_{z}$ and $\bAvgy\polDensityElem_{x}$
to the r.h.s.\ of Eqs.~(\ref{eq:BlsxTot}) and~(\ref{eq:BlszTot}),
resp. Then, we can write the Bloch equations as \begin{align}
\polDensityDotTot & =\left\langle \vec{b}\right\rangle \times\polDensityTot-\invTauSpinTensor\,\polDensityTot+\bGamma,\label{eq:BlVec}\end{align}
with $\bGamma=\big(\frac{1}{2}\nu\ZeemanX\,\tauSpinInPlane^{-1},\:\frac{1}{2}\nu\bAvgy\tauSpinInPlane^{-1},\:\frac{1}{4}\nu\bAvgy\ZeemanX\,\gammaHomTildeN\big)$.

\section{Physical quantities expressed in terms of Fourier coefficients}

\subsection{Transport lifetime}

The inverse transport lifetime is

\begin{equation}
\tau^{-1}=\int_{0}^{2\pi}K\left(\theta\right)(1-\cos\theta)=2\pi\left(K_{0}-K_{1}\right).\label{eq:tauToKFT}\end{equation}
This motivated our definition \begin{equation}
k_{m}=2\pi\tau\left(K_{0}-K_{m}\right)=\frac{K_{0}-K_{m}}{K_{0}-K_{1}},\label{eq:defkm}\end{equation}
and one can see that $k_{m}>0$ for $m\geq1$.

\subsection{Spin polarization density and spin currents}

\label{ssec:polarizationFT}We now evaluate the spin polarization
density $\polDensityTot\left(\vec{r}\right)=\sum_{\alpha\beta}\langle\psi_{\alpha}^{\dagger}\left(\vec{r}\right)\bsigma_{\alpha\beta}\psi_{\beta}\left(\vec{r}\right)\rangle$
and the spin current $\vec{j}^{\mu}\left(\vec{r}\right)=\sum_{\alpha\beta}\langle\psi_{\alpha}^{\dagger}\left(\vec{r}\right)\frac{1}{2}\,\{\sigma_{\alpha\beta}^{\mu},\:\vec{v}\}\psi_{\beta}\left(\vec{r}\right)\rangle$.
The spin polarization density can readily be expressed in terms of
the Fourier transformed distribution function, \begin{equation}
\polDensityElem_{\mu}=\iint d\epsilon d\azimuthalAngle\:\mathrm{Tr}\,\sigma_{\mu}\fMatrixE(\epsilon,\,\azimuthalAngle)=4\pi\int d\epsilon\:\Phimum{\mu}{m=0}.\label{eq:smuToFT}\end{equation}

For the spin current, one needs to evaluate \begin{equation}
\vec{j}^{\mu}=\iint d\epsilon d\theta\:\mathrm{Tr}\,\frac{1}{2}\,\{\sigma_{\mu},\:\vec{v}\}\fMatrixE(\epsilon,\,\azimuthalAngle),\label{eq:jspinToVSigmafMatrixE}\end{equation}
which is somewhat more complicated, as the velocity operator $\vec{v}$
depends on spin and on $\azimuthalAngle$. It is obtained from the
Heisenberg equation as \begin{equation}
\vec{v}=i\left[H,\,\vec{x}\right]=-\left[H,\,\partial_{\vec{k}}\right]=\frac{\partial\epsilon_{k}}{\partial\vec{k}}-\frac{1}{2}\,\frac{\partial\vec{b}\cdot\bsigma}{\partial\vec{k}}=\vec{v}_{0}+\delta\vec{v}.\label{eq:vTovnotAv}\end{equation}
For a fixed $\epsilon$, the value of the wave vector $k$ depends
on the spin, \begin{equation}
k=k_{\epsilon}+\frac{\bsigma\cdot\vec{b}}{2v_{\epsilon}}\label{eq:kTokepsilonSigmab}\end{equation}
{[}cf.\ Eq.~(B35) in \refcite{ShytovKinetic}{]}. Thus, for $\vec{v}_{0}=\partial\epsilon_{k}/\partial\vec{k}=\kUnitVec\, v_{0}$
and in lowest order in $\vec{b}$,\begin{align}
\frac{\partial v_{0}}{\partial\vec{b}} & =\frac{\partial v_{0}}{\partial k_{\epsilon}}\,\frac{\partial k_{\epsilon}}{\partial\vec{b}}\stackrel{(\ref{eq:kTokepsilonSigmab})}{=}\frac{\partial v_{0}}{\partial k_{\epsilon}}\,\frac{\bsigma}{2v_{\epsilon}}\approx\frac{\partial v_{\epsilon}}{\partial k_{\epsilon}}\,\frac{\bsigma}{2v_{\epsilon}}=\left(1+\zeta\right)\frac{\bsigma}{2k_{\epsilon}}.\label{eq:v0db}\end{align}
 The velocity $\vec{v}$ at a fixed energy $\epsilon$ can now be
expanded in $\vec{b}$, \begin{equation}
\vec{v}\stackrel{(\ref{eq:vTovnotAv}),(\ref{eq:v0db})}{=}\kUnitVec\:\left[v_{\epsilon}+\left(1+\zeta\right)\frac{\vec{b}\cdot\bsigma}{2k_{\epsilon}}\right]+\delta\vec{v},\label{eq:vnotTovepsilonDer}\end{equation}
yielding Eq.~(\ref{eq:vnotTovepsilon}). {[}Equation~(\ref{eq:vnotTovepsilon})
is consistent with the gradient term $\frac{1}{2}[\vec{v}\cdot\nabla\hat{f}+(\nabla\hat{f})\cdot\vec{v}]$
in Eq.~(B30) of \refcite{ShytovKinetic}.{]} Next, noting that $\frac{1}{2}\,\{\sigma_{\mu},\:\vec{b}\cdot\bsigma\}=b_{\mu}$,
we find \begin{align}
\frac{1}{2}\,\{\sigma_{\mu},\:\vec{v}\} & \stackrel{(\ref{eq:vnotTovepsilon})}{=}\sigma_{\mu}\kUnitVec v_{\epsilon}+\kUnitVec\,\left(1+\zeta\right)\frac{b_{\mu}}{2k_{\epsilon}}-\frac{1}{2}\,\frac{\partial b_{\mu}}{\partial\vec{k}};\label{eq:antiCommutsigmav}\end{align}
and we see that the second term in Eq.~(\ref{eq:antiCommutsigmav})
is not present in \refcite{ShytovKinetic}, but would not lead to
an extra contribution to $\vec{j}^{\mu}$ there. 

We insert Eq.~(\ref{eq:antiCommutsigmav}) into Eq.~(\ref{eq:jspinToVSigmafMatrixE})
and use $\fMatrixE=\frac{1}{2}\fDiagonalE+\bsigma\cdot\fSpinE$, thus
\begin{align}
\vec{j}^{\mu} & =\iint d\epsilon d\theta\:\left[2\kUnitVec v_{\epsilon}\fSpinEComponent{\mu}+\kUnitVec\,\left(1+\zeta\right)\frac{b_{\mu}}{2k_{\epsilon}}\fDiagonalE-\frac{1}{2}\,\frac{\partial b_{\mu}}{\partial\vec{k}}\fDiagonalE\right]\label{eq:jspinTofMatrixE}\\
 & =\vec{j}^{\jSpLblPhimu\mu}+\vec{j}^{\jSpLblPhiC\mu}.\end{align}

The components of the first term in Eq.~(\ref{eq:jspinTofMatrixE})
are \begin{align}
j_{x}^{\jSpLblPhimu\mu} & =2\iint d\epsilon d\theta\: v_{\epsilon}\cos\theta\:\fSpinEComponent{\mu}(\epsilon,\,\theta)\\
 & =2\pi\int d\epsilon\: v_{\epsilon}(\Phimum{\mu}{1}+\Phimum{\mu}{-1}),\label{eq:SC1muxToFT}\\
j_{y}^{\jSpLblPhimu\mu} & =2\iint d\epsilon d\theta\: v_{\epsilon}\sin\theta\:\fSpinEComponent{\mu}(\epsilon,\,\theta)\\
 & =2\pi i\int d\epsilon\: v_{\epsilon}(\Phimum{\mu}{1}-\Phimum{\mu}{-1}).\label{eq:SC1muyToFT}\end{align}
We evaluate Eqs.~(\ref{eq:SC1muxToFT}) and~(\ref{eq:SC1muxToFT})
for concrete $\mu$, use $\PhiPMm{m}=\Phimum{x}{m}+i\Phimum{y}{m}$,
and take advantage of the fact that the spin current is a real quantity.
This yields\begin{align}
j_{x}^{\jSpLblPhimu z} & =4\pi\int d\epsilon\: v_{\epsilon}\mathrm{Re}\fSpinEComponent{z}_{1},\label{eq:SCzxToFT}\\
j_{y}^{\jSpLblPhimu z} & =-4\pi\int d\epsilon\: v_{\epsilon}\mathrm{Im}\fSpinEComponent{z}_{1},\\
j_{x}^{\jSpLblPhimu x} & =2\pi\int d\epsilon\: v_{\epsilon}\,\mathrm{Re}\left[\Phi_{1}+\Phi_{-1}\right],\label{eq:SCxxToFT}\\
j_{y}^{\jSpLblPhimu x} & =-2\pi\int d\epsilon\: v_{\epsilon}\,\mathrm{Im}\left[\Phi_{1}-\Phi_{-1}\right],\\
j_{x}^{\jSpLblPhimu y} & =2\pi\int d\epsilon\: v_{\epsilon}\,\mathrm{Im}\left[\Phi_{1}+\Phi_{-1}\right],\\
j_{y}^{\jSpLblPhimu y} & =2\pi\int d\epsilon\: v_{\epsilon}\,\mathrm{Re}\left[\Phi_{1}-\Phi_{-1}\right].\label{eq:SCyxToFT}\end{align}

\begin{widetext}Finally, we consider the remaining terms in Eq.~(\ref{eq:jspinTofMatrixE}).
Because $\azimuthalAngle=\arctan(k_{y}/k_{x})$, we have $\partial\azimuthalAngle/\partial k_{x}=-\sin(\azimuthalAngle)/k$
and $\partial\azimuthalAngle/\partial k_{y}=\cos(\azimuthalAngle)/k$,
thus \begin{align}
\frac{\partial b_{\mu}}{\partial k_{x}} & =\frac{\partial k}{\partial k_{x}}\,\frac{\partial b_{\mu}}{\partial k}+\frac{\partial\azimuthalAngle}{\partial k_{x}}\,\frac{\partial b_{\mu}}{\partial\azimuthalAngle}=\cos\azimuthalAngle\,\frac{\partial b_{\mu}}{\partial k}-\frac{\sin\azimuthalAngle}{k_{\epsilon}}\,\frac{\partial b_{\mu}}{\partial\azimuthalAngle},\\
\frac{\partial b_{\mu}}{\partial k_{y}} & =\frac{\partial k}{\partial k_{y}}\,\frac{\partial b_{\mu}}{\partial k}+\frac{\partial\azimuthalAngle}{\partial k_{y}}\,\frac{\partial b_{\mu}}{\partial\azimuthalAngle}=\sin\azimuthalAngle\,\frac{\partial b_{\mu}}{\partial k}+\frac{\cos\azimuthalAngle}{k_{\epsilon}}\,\frac{\partial b_{\mu}}{\partial\azimuthalAngle}.\end{align}
We explicitly evaluate $\vec{j}^{\jSpLblPhiC\mu}$ by inserting $\fDiagonalE=\frac{eE\tau}{2\pi^{2}}\left(\frac{-\partial f_{0}}{\partial\epsilon}\right)k_{\epsilon}\cos\azimuthalAngle$
into Eq.~(\ref{eq:jspinTofMatrixE}), \begin{align}
j_{x}^{\jSpLblPhiC\mu} & =\iint d\epsilon d\theta\:\frac{eE\tau}{4\pi^{2}}\left(\frac{-\partial f_{0}}{\partial\epsilon}\right)\left[\cos^{2}\azimuthalAngle\,\left(1+\zeta\right)b_{\mu}-k_{\epsilon}\cos^{2}\azimuthalAngle\,\frac{\partial b_{\mu}}{\partial k}+\sin\azimuthalAngle\cos\azimuthalAngle\,\frac{\partial b_{\mu}}{\partial\azimuthalAngle}\right]\\
 & =\int d\epsilon\:\frac{eE\tau}{8\pi}\left(\frac{-\partial f_{0}}{\partial\epsilon}\right)\left[\left(1+\zeta-k_{\epsilon}\frac{\partial}{\partial k}\right)\left(2\bFTComp{\mu}{0}+\bFTComp{\mu}{2}+\bFTComp{\mu}{-2}\right)-2(\bFTComp{\mu}{2}+\bFTComp{\mu}{-2})\right],\label{eq:SC2muxToFT}\\
j_{y}^{\jSpLblPhiC\mu} & =\iint d\epsilon d\theta\:\frac{eE\tau}{4\pi^{2}}\left(\frac{-\partial f_{0}}{\partial\epsilon}\right)\left[\sin\azimuthalAngle\cos\azimuthalAngle\,\left(1+\zeta\right)b_{\mu}-k_{\epsilon}\sin\azimuthalAngle\cos\azimuthalAngle\,\frac{\partial b_{\mu}}{\partial k}-\cos^{2}\azimuthalAngle\frac{\partial b_{\mu}}{\partial\azimuthalAngle}\right]\\
 & =\int d\epsilon\:\frac{eE\tau}{8\pi}\left(\frac{-\partial f_{0}}{\partial\epsilon}\right)\: i\left(-1+\zeta-k_{\epsilon}\frac{\partial}{\partial k}\right)\left(\bFTComp{\mu}{2}-\bFTComp{\mu}{-2}\right),\label{eq:SC2muyToFT}\end{align}
where we decomposed $b_{\mu}=\sum_{m}\, e^{im\azimuthalAngle}\bFTComp{\mu}{m}$
and used $\cos^{2}\azimuthalAngle=\frac{1}{4}\left(2+e^{-2i\azimuthalAngle}+e^{2i\azimuthalAngle}\right)$
and $\sin\azimuthalAngle\cos\azimuthalAngle=\frac{i}{4}\left(e^{-2i\azimuthalAngle}-e^{2i\azimuthalAngle}\right)$.

Thus, for total spin currents $\vec{j}^{\mu}$, we find \begin{align}
j_{x}^{\mu} & \stackrel{(\ref{eq:SC1muxToFT}),(\ref{eq:SC2muxToFT})}{=}2\pi\int d\epsilon\: v_{\epsilon}\left\{ \Phimum{\mu}{1}+\Phimum{\mu}{-1}+\Eparam\tau\left[\left(1+\zeta-k_{\epsilon}\frac{\partial}{\partial k}\right)\left(2\bFTComp{\mu}{0}+\bFTComp{\mu}{2}+\bFTComp{\mu}{-2}\right)-2(\bFTComp{\mu}{2}+\bFTComp{\mu}{-2})\right]\right\} ,\\
j_{y}^{\mu} & \stackrel{(\ref{eq:SC1muyToFT}),(\ref{eq:SC2muyToFT})}{=}2\pi i\int d\epsilon\: v_{\epsilon}\left\{ \Phimum{\mu}{1}-\Phimum{\mu}{-1}+\Eparam\tau\left(-1+\zeta-k_{\epsilon}\frac{\partial}{\partial k}\right)\left(\bFTComp{\mu}{2}-\bFTComp{\mu}{-2}\right)\right\} .\end{align}
For the model of Eq.~(\ref{eq:bofk}), we then use $\bFTComp{\mu}{2}=\bFTComp{\mu}{-2}=0$
and $\bFTComp{\mu}{0}=\ZeemanX\delta_{\mu x}$.\vspace{5mm}\end{widetext}

For comparison, note that the charge current is \begin{align}
\vec{j}^{\mathrm{c}} & =\iint d\epsilon d\theta\: e\vec{v}\:\mathrm{Tr}\,\fMatrixE(\epsilon,\theta)=\int\int d\epsilon d\theta\: e\vec{v}\fDiagonalE.\end{align}
Evaluating the distribution $\fDiagonalE$, we get \begin{align}
j_{x}^{\mathrm{c}} & =\iint d\epsilon d\theta\:8e\Eparam\tau v_{\epsilon}^{2}k_{\epsilon}\cos^{2}\theta=\int d\epsilon\:8\pi e\Eparam\tau v_{\epsilon}^{2}k_{\epsilon}\label{eq:jCxToEparam}\\
 & \stackrel{T=0,\,\zeta=0}{=}e^{2}E\tau\,\frac{v_{\mathrm{F}}k_{\mathrm{F}}}{2\pi}=\frac{e^{2}E\tau}{m^{*}}\,\frac{k_{\mathrm{F}}^{2}}{2\pi}=\frac{e^{2}E\tau\nTwoD}{m^{*}},\end{align}
i.e., we recover the Drude conductivity.

\subsection{Expressing $\tilde{K}_{m}$ in terms of $K_{n}$}

\label{sub:tildeKtoK}Using that $K\left(\theta\right)=\sum_{n=-\infty}^{\infty}e^{in\theta}\, K_{n}$
and that $K_{-m}=K_{m}$, we get \begin{align}
\tilde{K}\left(\theta\right) & =\tan\frac{\theta}{2}\:\frac{\partial K\left(\theta\right)}{\partial\theta}\\
 & =\tan\frac{\theta}{2}\:\sum_{n>0}\, in\,\left(e^{in\theta}-e^{-in\theta}\right)\, K_{n}\\
 & =-2\sum_{n>0}nK_{n}\,\sin\left(n\theta\right)\,\tan\frac{\theta}{2}.\end{align}
Because $\tilde{K}\left(-\theta\right)=\tilde{K}\left(\theta\right)$,
we get \begin{align}
\tilde{K}_{m} & =\frac{1}{2\pi}\int d\theta\:\frac{e^{-im\theta}+e^{im\theta}}{2}\,\tilde{K}\left(\theta\right)\\
 & =-2\sum_{n>0}nK_{n}\;\frac{1}{2\pi}\int d\theta\:\cos\left(m\theta\right)\,\sin\left(n\theta\right)\,\tan\frac{\theta}{2}.\end{align}
For $n>0$, we integrate  \begin{align}
I_{m,n} & =\frac{1}{2\pi}\int d\theta\:\cos\left(m\theta\right)\,\sin\left(n\theta\right)\,\tan\frac{\theta}{2}\\
 & =\begin{cases}
0 & n<\left|m\right|,\\
-\frac{1}{2} & n=\left|m\right|,\\
\left(-1\right)^{1+n+m} & n>\left|m\right|,\end{cases}\end{align}
 which leads to \begin{align}
\tilde{K}_{m} & =\left|m\right|K_{m}+2\left(-1\right)^{m}\sum_{n>\left|m\right|}\left(-1\right)^{n}nK_{n}\label{eq:KTildeToK}\end{align}
 Thus, we can use\begin{equation}
2\!\sum_{n>\left|m\right|}\left(-1\right)^{n}nK_{n}=\tilde{K}_{0}-2\!\sum_{0<n\leq\left|m\right|}\left(-1\right)^{n}nK_{n},\end{equation}
and find\begin{align}
\tilde{K}_{m} & =\left(-1\right)^{m}\tilde{K}_{0}-\left|m\right|K_{m}-2\left(-1\right)^{m}\sum_{0<n<\left|m\right|}\left(-1\right)^{n}nK_{n}.\label{eq:KtildeMToKtildeNotKm}\end{align}
 .\clearpage
\end{document}